\documentclass{article}
\pdfoutput=1 
\usepackage{url}
\usepackage[colorlinks=true]{hyperref}
\usepackage{spconf}
\usepackage{color,xcolor}
\usepackage{epsfig}
\usepackage{graphicx}
\usepackage{array}
\usepackage{booktabs}
\usepackage{colortbl}
\usepackage{multirow}
\usepackage{subcaption}
\usepackage{amsmath,amsfonts,amsthm,amssymb,bm}
\usepackage{changepage}
\usepackage{extramarks}
\usepackage{fancyhdr}
\usepackage{lastpage}
\usepackage{setspace}
\usepackage{soul}
\usepackage{xspace}
\usepackage{algorithm, algorithmic}
\usepackage{enumitem}

\newlength{\bibitemsep}
\newlength{\bibparskip}
\let\oldthebibliography\thebibliography
\renewcommand\thebibliography[1]{%
  \oldthebibliography{#1}%
  \setlength{\parskip}{\bibitemsep}%
  \setlength{\itemsep}{\bibparskip}%
}
\newcommand{\sect}[1]{Section~\ref{sec:#1}}

\newcommand{\eqn}[1]{Eqn.~\ref{eqn:#1}}
\newcommand{\fig}[1]{Fig.~\ref{fig:#1}}
\newcommand{\tab}[1]{Table~\ref{tab:#1}}

\newcommand{\ignore}[1]{}

\makeatletter
\DeclareRobustCommand\onedot{\futurelet\@let@token\@onedot}
\def\@onedot{\ifx\@let@token.\else.\null\fi\xspace}
 
\def\ie{\emph{i.e}\onedot}

\makeatother
\def\presec{\vspace{-0.5em}}
\def\postsec{\vspace{-0.5em}}

\title{Bi-Directional Lattice Recurrent Neural Networks\\for Confidence Estimation}
\name{Q.~Li$^\dagger$, P.~M.~Ness$^\dagger$, A.~Ragni$^\ddagger$, M.~J.~F. Gales$^\ddagger$}
\address{Department of Engineering, University of Cambridge\\Trumpington Street, Cambridge CB2 1PZ, UK\\\texttt{\small{\{ql264, pmn26, ar527, mjfg\}@eng.cam.ac.uk}}\thanks{$^\dagger$ Both authors contributed equally. P.~M. Ness was supported in part by the ALTA Institute, Cambridge University. $^\ddagger$ Supported in part by the Office of the Director of National Intelligence (ODNI), Intelligence Advanced Research Projects Activity (IARPA), via Air Force Research Laboratory (AFRL) contract \# FA8650-17-C-9117. The views and conclusions contained herein are those of the authors and should not be interpreted as necessarily representing the official policies, either expressed or implied, of ODNI, IARPA, AFRL or the U.S. Government. The U.S. Government is authorised to reproduce and distribute reprints for governmental purposes notwithstanding any copyright annotation therein.
}}

\begin{document}
\ninept
\maketitle

\begin{abstract}
The standard approach to mitigate errors made by an automatic speech recognition system is to use confidence scores associated with each predicted word. In the simplest case, these scores are word posterior probabilities whilst more complex schemes utilise bi-directional recurrent neural network (BiRNN) models. A number of upstream and downstream applications, however, rely on confidence scores assigned not only to 1-best hypotheses but to all words found in confusion networks or lattices. These include but are not limited to speaker adaptation, semi-supervised training and information retrieval. Although word posteriors could be used in those applications as confidence scores, they are known to have reliability issues. To make improved confidence scores more generally available, this paper shows how BiRNNs can be extended from 1-best sequences to confusion network and lattice structures. Experiments are conducted using one of the Cambridge University submissions to the IARPA OpenKWS 2016 competition. The results show that confusion network and lattice-based BiRNNs can provide a significant improvement in confidence estimation.
\end{abstract}

\begin{keywords}
confidence estimation, bi-directional recurrent neural network, confusion network, lattice
\end{keywords}

\section{Introduction}
\postsec
\label{sec:intro}
Recent years have seen an increased usage of spoken language technology in applications ranging from speech transcription~\cite{Xiong2018TheM2} to personal assistants~\cite{Li2017AcousticMF}. The quality of these applications heavily depends on the accuracy of the underlying automatic speech recognition (ASR) system yielding 1-best hypotheses and how well ASR errors are mitigated. The standard approach to ASR error mitigation is confidence scores~\cite{Wessel2001ConfidenceMF, Jiang2005ConfidenceMF}. A low confidence can give a signal to downstream applications about the high uncertainty of the ASR in its prediction and measures can be taken to mitigate the risk of making a wrong decision. However, confidence scores can also be used in upstream applications such as speaker adaptation~\cite{Uebel2001SpeakerAU} and semi-supervised training~\cite{Chan2004ImprovingBN, Tr2005CombiningAA} to reflect uncertainty among {\em multiple} possible alternative hypotheses. Downstream applications, such as machine translation and information retrieval, could similarly benefit from using multiple hypotheses.

A range of confidence scores has been proposed in the literature~\cite{Jiang2005ConfidenceMF}. In the simplest case, confidence scores are posterior probabilities that can be derived using approaches such as confusion networks~\cite{Mangu2000FindingCI, Evermann2000LargeVD}. These posteriors typically significantly over-estimate confidence~\cite{Evermann2000LargeVD}. Therefore, a number of approaches have been proposed to rectify this problem. These range from simple piece-wise linear mappings given by decision trees~\cite{Evermann2000LargeVD} to more complex sequence models such as conditional random fields~\cite{Seigel2011CombiningIS}, and to neural networks~\cite{Kalgaonkar2015EstimatingCS, DelAgua2018SpeakerAdaptedCM, Ragni2018ConfidenceEA}. Though improvements over posterior probabilities on 1-best hypotheses were reported, the impact of these approaches on all hypotheses available within confusion networks and lattices has not been investigated. 

Extending confidence estimation to confusion network and lattice structures can be straightforward for some approaches, such as decision trees, and challenging for others, such as recurrent forms of neural networks.  The previous work on encoding graph structures into neural networks~\cite{Scarselli2009TheGN} has mostly focused on embedding lattices into a fixed dimensional vector representation~\cite{Su2017LatticeBasedRN, Ladhak2016LatticeRnnRN}. This paper examines a particular example of extending a bi-directional recurrent neural network (BiRNN)~\cite{Schuster1997BidirectionalRN} to confusion network and lattice structures. This requires specifying how BiRNN states are propagated in the forward and backward directions, how to merge a variable number of BiRNN states, and how target confidence values are assigned to confusion network and lattice arcs. The paper shows that the state propagation in the forward and backward directions has close links to the standard forward-backward algorithm~\cite{Rabiner1989tutorial}. This paper proposes several approaches for merging BiRNN states, including an attention mechanism \cite{Vaswani2017AttentionIA}. Finally, it describes a Levenshtein algorithm for assigning targets to confusion networks and an approximate solution for lattices. Combined these make it possible to assign confidence scores to every word hypothesised by the ASR, not just from a single extracted hypothesis. 

The rest of this paper is organised as follows. \sect{recurrent} describes the use of bi-directional recurrent neural networks for confidence estimation in 1-best hypotheses. \sect{recursive} describes the extension to confusion network and lattice structures. Experimental results are presented in \sect{exp}. The conclusions drawn from this work are given in \sect{conclusion}.

\presec
\section{Bi-Directional Recurrent Neural Network}
\begin{figure}[ht]
    \centering
    \begin{subfigure}[t]{0.3\linewidth}
        \centering
        \includegraphics[height=4cm]{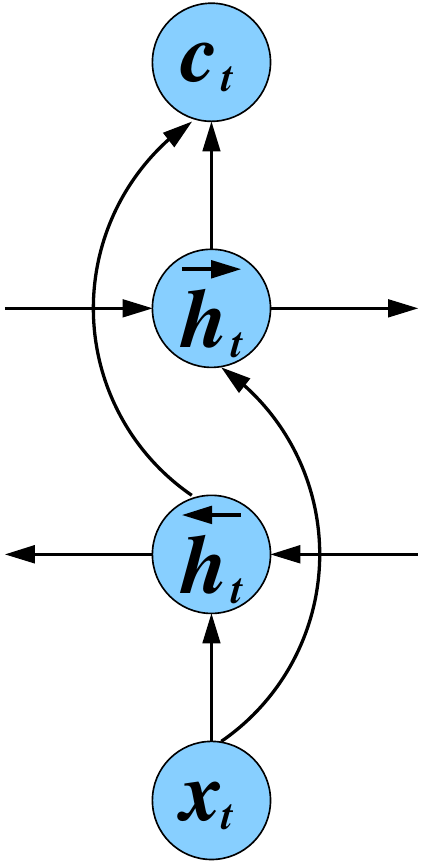}
        \caption{sequence}
        \label{fig:recurrent}
    \end{subfigure}
    \begin{subfigure}[t]{0.45\linewidth}
        \centering
        \includegraphics[height=4cm]{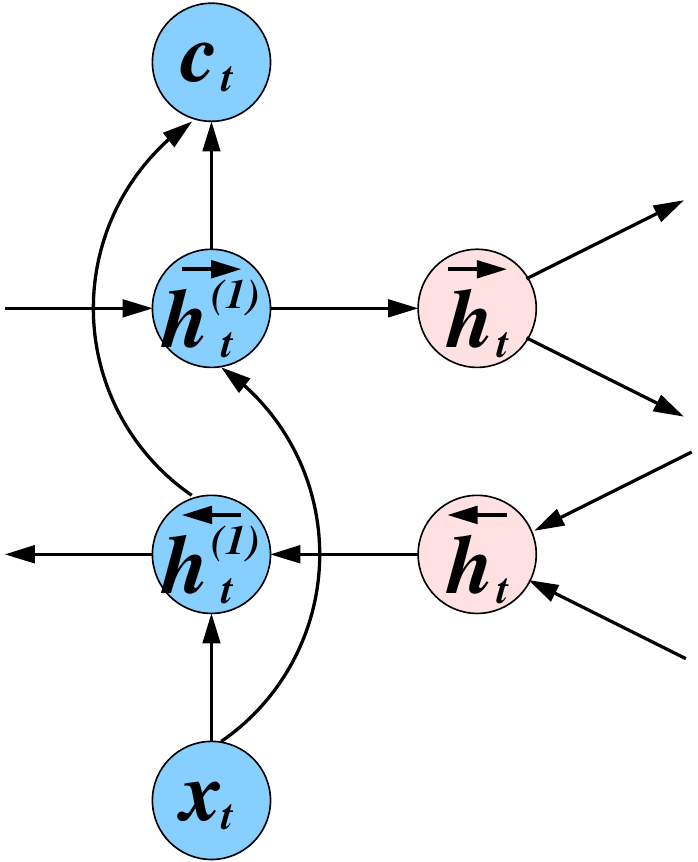}
        \caption{confusion network, lattice}
        \label{fig:recursive}
    \end{subfigure}
    \caption{Bi-directional neural networks for confidence estimation}
    \vspace{-1em}
\end{figure}

\postsec
\label{sec:recurrent}
\fig{recurrent} shows the simplest form of the BiRNN~\cite{Schuster1997BidirectionalRN}. Unlike its uni-directional version, the BiRNN makes use of two recurrent states, one going in the forward direction in time $\overrightarrow{\mathbf{h}}_{t}$ and another in the backward direction $\overleftarrow{\mathbf{h}}_{t}$ to model past (history) and future information respectively. 
The past information can be modelled by
\begin{equation}
    \overrightarrow{\mathbf{h}}_{t} = \sigma(\mathbf{ W}^{(\overrightarrow{{h}})}\overrightarrow{\mathbf{h}}_{t-1} + \mathbf{ W}^{(x)}\mathbf{x}_{t})
    \label{eqn:rnn_update}
\end{equation}
where $\mathbf{x}_{t}$ is an input feature vector at time $t$, $\mathbf{W}^{(x)}$ is an input matrix, $\mathbf{W}^{(\overrightarrow{{h}})}$ is a history matrix and $\sigma$ is an element-wise non-linearity such as a sigmoid. The future information is typically modelled in the same way. At any time $t$ the confidence $c_t$ can be estimated by
\begin{equation}
    c_{t} = \sigma(\mathbf{w}^{(c)^{\sf T}}{\bf h}_{t} + {b}^{(c)})
    \label{eqn:fc}
\end{equation}
where $\mathbf{w}^{c}$ and $b^{(b)}$ are a parameter vector and a bias, $\sigma$ is any non-linearity that maps confidence score into the range $[0,1]$ and $\mathbf{h}_{t}$ is a context vector that combines the past and future information.
\begin{equation}
    \mathbf{h}_{t} = \begin{bmatrix}\overrightarrow{\bf h}_{t} & \overleftarrow{\bf h}_{t}\end{bmatrix}^{\sf T}
    \label{eqn:concat}
\end{equation}
The input features $\mathbf{x}_{t}$ play a fundamental role in the model's ability to assign accurate confidence scores. Numerous hand-crafted features have been proposed~\cite{Schaaf1997ConfidenceMF, Weintraub1997NeuralBM, Ma2007UnsupervisedTO, Seigel2014DetectingDI}. In the simplest case, duration and word posterior probability can be used as input features. More complex features may include embeddings~\cite{Mikolov2013DistributedRO}, acoustic and language model scores and other information. The BiRNN can be trained by minimising the binary cross-entropy 
\begin{equation}
    H(\mathbf{c},\mathbf{c}^{*};\bm{\theta}) = -\dfrac{1}{T}\sum_{t=1}^{T} \Big\{{c}_{t}^{*} \log(c_{t}) + (1 - {c}_{t}^{*}) \log(1 - c_{t})\Big\}
    \label{eqn:ce}
\end{equation}
where $c_{t}$ is a predicted confidence score for time slot $t$ and $c_{t}^{*}$ is the associated reference value. The reference values can be obtained by aligning the 1-best ASR output and reference text using the Levenshtein algorithm. Note that deletion errors cannot be handled under this framework and need to be treated separately~\cite{Seigel2014DetectingDI, Ragni2018ConfidenceEA}. This form of BiRNN has been examined for confidence estimation in~\cite{DelAgua2018SpeakerAdaptedCM, Ragni2018ConfidenceEA}

The perfect confidence estimator would assign scores of one and zero to correctly and incorrectly hypothesised words respectively. In order to measure the accuracy of confidence predictions, a range of metrics have been proposed. Among these, normalised cross-entropy (NCE) is the most frequently used ~\cite{siu1997improved}. NCE measures the relative change in the binary cross-entropy when the empirical estimate of ASR correctness, $P_c$, is replaced by predicted confidences $\mathbf{c}={c_1,\ldots,c_T}$. Using the definition of binary cross-entropy in \eqn{ce}, NCE can be expressed as
\begin{equation}
    \text{NCE}(\mathbf{c},\mathbf{c^*}) = 
    \dfrac{H(P_{c}\cdot\textbf{1},\mathbf{c^*}) - H(\mathbf{c},\mathbf{c^*})}{H(P_{c}\cdot\textbf{1},\mathbf{c^*})}
    \label{eqn:NCE}
\end{equation}
where $\mathbf{1}$ is a length $T$ vector of ones, and the empirical estimate of ASR correctness is given by
\begin{equation}
    P_{c} = \dfrac{1}{T}\sum_{t=1}^{T} {c}_{t}^{*}
    \label{eqn:p_c}
\end{equation}
When hypothesised confidence scores $\mathbf{c}$ are systematically better than the estimate of ASR correctness $P_c$, NCE is positive. In the limit of perfect confidence scores, NCE approaches one.

NCE alone is not always the most optimal metric for evaluating confidence estimators. This is because the theoretical limit of correct words being assigned a score of one and incorrect words a score of zero is not necessary for perfect operation of an upstream or downstream application. Often it is sufficient that the rank ordering of the predictions is such that all incorrect words fall below a certain threshold, and all correct words above. This is the case, for instance, in various information retrieval tasks~\cite{Gales2017aL, Ragni2018AutomaticSR}. A more suitable metric in such cases could be an area under a curve (AUC)-type metric. For balanced data the chosen curve is often the receiver operation characteristics (ROC). Whereas for imbalanced data, as is the case in this work, the precision-recall (PR) curve is normally used~\cite{ROCvPRcurves}. The PR curve is obtained by plotting precision versus recall 
\begin{equation}
    \text{Precision}(\theta) = \dfrac{\text{TP}(\theta)}{\text{TP}(\theta)+\text{FP}(\theta)},\;
    \text{Recall}(\theta) = \dfrac{\text{TP}(\theta)}{\text{TP}(\theta) + \text{FN}(\theta)}
    \label{eqn:PR}
\end{equation}
for a range of thresholds $\theta$, where TP are true positives, FP and FN are false positives and negatives. When evaluating performance on lattices and confusion networks, these metrics are computed across all arcs in the network.

\presec
\section{Confusion Network and Lattice Extensions}
\begin{figure}[ht]
    \centering
    \begin{subfigure}{.48\linewidth}
        \centering
        \vspace{-7em}
        \includegraphics[width=0.95\linewidth]{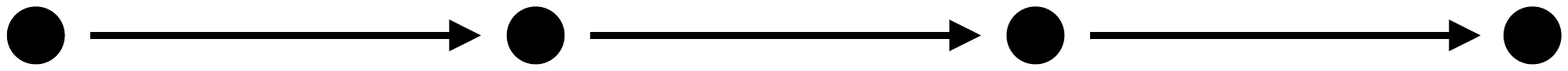}
        \vspace{-8em}
        \caption{one-best sequence}
        \label{fig:one-best}
        \vspace{-6.5em}
        \includegraphics[width=0.95\linewidth]{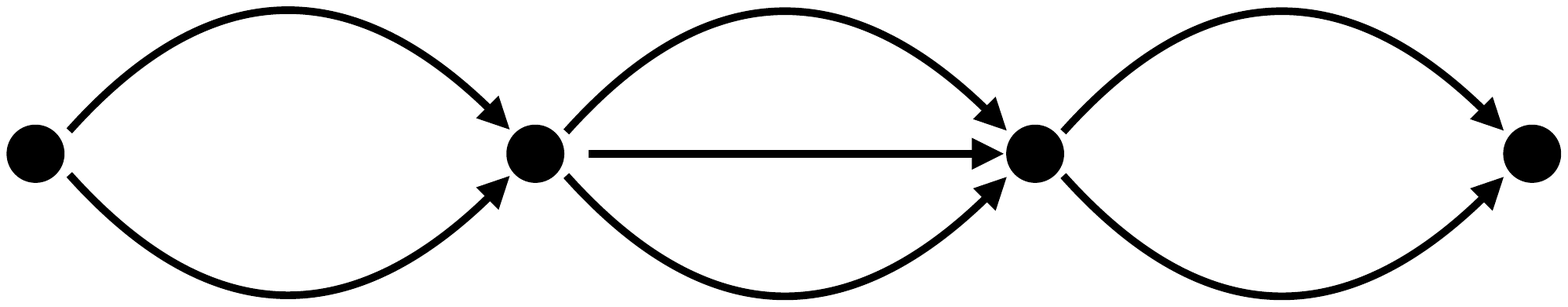}
        \vspace{-7.5em}
        \caption{confusion network}
        \label{fig:cn}
    \end{subfigure}
    \hfill
    \begin{subfigure}{0.48\linewidth}
        \centering
        \vspace{-5em}
        \includegraphics[width=\linewidth]{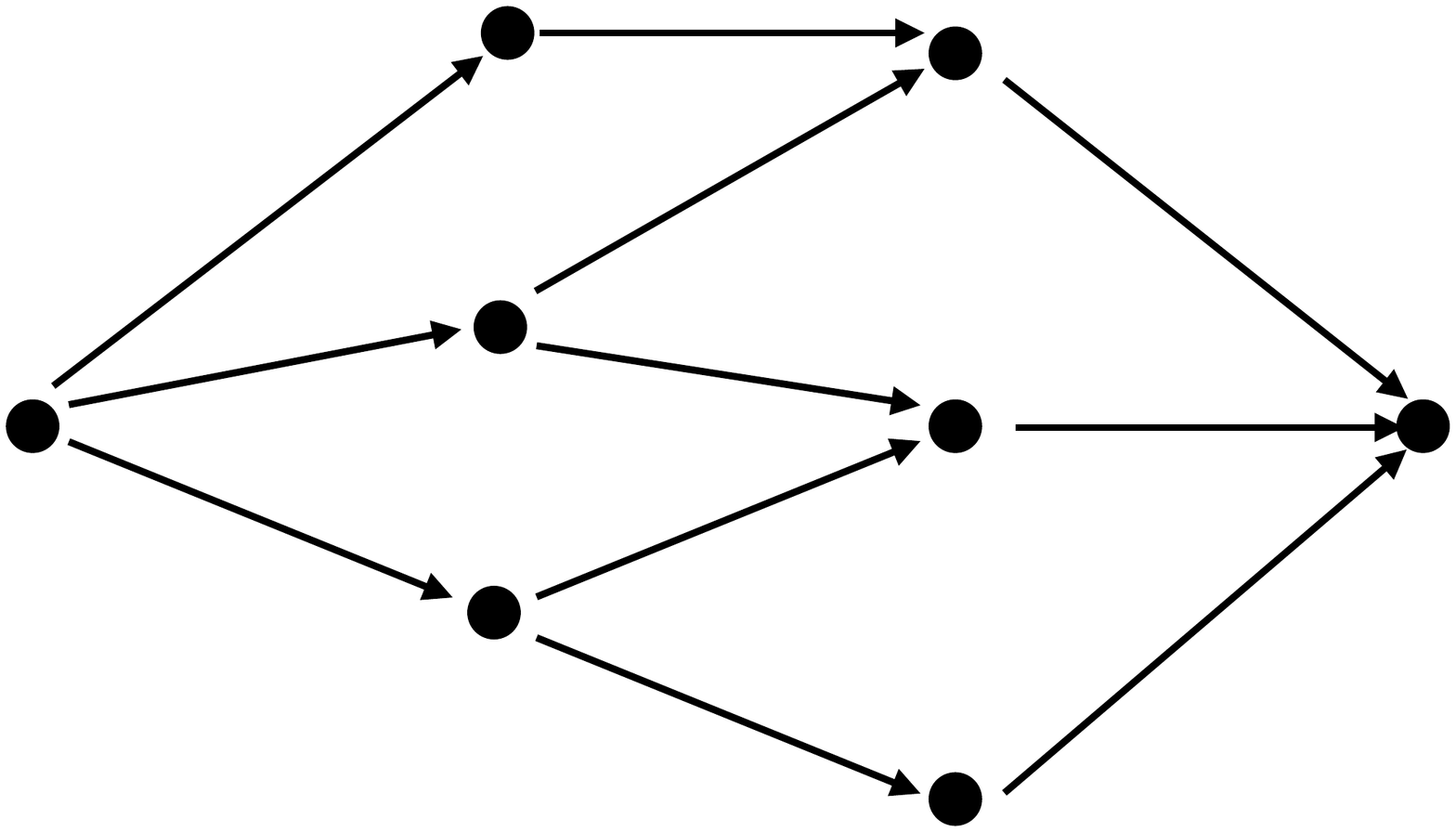}
        \vspace{-6.5em}
        \caption{word lattice}
        \label{fig:lattice}
    \end{subfigure}
    \caption{Standard ASR outputs}
    \label{fig:structures}
    \vspace{-1em}
\end{figure}

\postsec
\label{sec:recursive}
A number of important downstream and upstream applications rely on accurate confidence scores in graph-like structures, such as confusion networks (CN) in \fig{cn} and lattices in \fig{lattice}, where arcs connected by nodes represent hypothesised words. This section describes an extension of BiRNNs to CNs and lattices.

\fig{cn} shows that compared to 1-best sequences in \fig{one-best}, each node in a CN may have multiple incoming arcs. Thus, a decision needs to be made on how to optimally propagate information to the outgoing arcs. Furthermore, any such approach would need to handle a variable number of incoming arcs. One popular approach~\cite{Ladhak2016LatticeRnnRN, Su2017LatticeBasedRN} is to use a weighted combination
\begin{equation}
    \overrightarrow{\mathbf{h}}_{t} = \sum_{i} \alpha_{t}^{(i)} \overrightarrow{\mathbf{h}}_{t}^{(i)}
    \label{eqn:forward_node_state}
\end{equation}
where $\overrightarrow{\mathbf{h}}_{t}^{(i)}$ represents the history information associated with the $i^{\text{th}}$ arc of the $t^{\text{th}}$ CN bin and $\alpha_{t}^{(i)}$ is the associated weight. A number of approaches can be used to set these weights. One simple approach is to set weights of all arcs other than the one with the highest posterior to zero. This yields a model that for 1-best hypotheses has no advantage over BiRNNs in \sect{recurrent}. Other simple approaches include average or normalised confidence score $\alpha_t^{(i)} = c_t^{(i)}/\sum_{j} c_t^{(j)}$ where $c_{t}^{(i)}$ is a word posterior probability, possibly mapped by decision trees. A more complex approach is an attention mechanism
\begin{equation}
    \alpha_{t}^{(i)} = \dfrac{\exp(z_{t}^{(i)})}{\sum_{j} \exp(z_{t}^{(j)})}, \;\text{where } z_{t}^{(i)} = \sigma\left({\mathbf{w}^{(a)}}^{\sf{T}}\overrightarrow{\mathbf{k}}_{t}^{(i)} + b^{(a)}\right)
\end{equation}
where $\mathbf{w}^{(a)}$ and $b^{(a)}$ are attention parameters, $\overrightarrow{\mathbf{k}}_{t}^{(i)}$ is a key. The choice of the key is important as it helps the attention mechanism decide which information should be propagated. It is not obvious a priori what the key should contain. One option is to include arc history information as well as some basic confidence score statistics
\begin{equation}
    \overrightarrow{\mathbf{k}}_{t}^{(i)} = \begin{bmatrix}
    \overrightarrow{\mathbf{h}}_{t}^{(i)^{\sf T}} & c_{t}^{(i)}  & \mu_{t} & \sigma_{t} \end{bmatrix}^{\sf T}
\end{equation}
where $\mu_t$ and $\sigma_t$ are the mean and standard deviation computed over $c_t^{(i)}$ at time $t$.
At the next $(t+1)^{\text{th}}$ CN bin the forward information associated with the $i^{\text{th}}$ arc is updated by 
\begin{equation}
    \overrightarrow{\mathbf{h}}_{t+1}^{(i)} = \sigma(\mathbf{ W}^{(\overrightarrow{{h}})}\overrightarrow{\mathbf{h}}_{t} + \mathbf{ W}^{(x)}\mathbf{x}_{t+1}^{(i)})
    \label{eqn:h_update}
 \end{equation}
The confidence score for each CN arc is computed by
\begin{equation}
c_{t}^{(i)} = \sigma(\mathbf{w}^{(c)^{\sf T}}{\bf h}_{t}^{(i)} + {b}^{(c)})
\end{equation}
where ${\bf h}_{t}^{(i)}$ is an arc context vector
\begin{equation}
    {\bf h}_{t}^{(i)} = \begin{bmatrix}
    \overrightarrow{\mathbf{h}}_{t}^{(i)} & \overleftarrow{\mathbf{h}}_{t}^{(i)}
    \end{bmatrix}
\end{equation}
A summary of dependencies in this model is shown in \fig{recursive} for a CN with 1 arc in the $t^{\text{th}}$ bin and 2 arcs in the $(t+1)^{\text{th}}$ bin.

As illustrated in \fig{lattice}, each node in a lattice marks a timestamp in an utterance and each arc represents a hypothesised word with its corresponding acoustic and language model scores. Although lattices do not normally obey a linear graph structure, if they are traversed in the topological order, no changes are required to compute confidences over lattice structures. The way the information is propagated in these graph structures is similar to the forward-backward algorithm~\cite{Rabiner1989tutorial}. There, the forward probability at time $t$ is
\begin{equation}
    \overrightarrow{h}_{t+1}^{(i)} = \overrightarrow{h}_{t} x_{t+1}^{(i)}, \;\text{where } \overrightarrow{h}_{t} = \sum_{j} \alpha_{i,j} \overrightarrow{h}_{t}^{(j)}
\end{equation}
Compared to equations \eqn{forward_node_state} and \eqn{h_update}, the forward recursion employs a different way to combine features $x_{t+1}^{(i)}$ and node states $\overrightarrow{h}_{t}$, and maintains stationary weights, \ie the transition probabilities $\alpha_{i,j}$, for combining arc states $\overrightarrow{h}_{t}^{(j)}$. In addition, each $\overrightarrow{h}_{t}^{(i)}$ has a probabilistic meaning which the vector $\overrightarrow{\mathbf{h}}_{t}^{(i)}$ does not. Furthermore, unlike in the standard algorithm, the past information at the final node is not constrained to be equal to the future information at the initial node.

In order to train these models, each arc of a CN or lattice needs to be assigned an appropriate reference confidence value. For aligning a reference word sequence to another sequence, the Levenshtein algorithm can be used. The ROVER method has been used to iteratively align word sequences to a pivot reference sequence to construct CNs~\cite{ROVER}. This approach can be extended to confusion network combination (CNC), which allows the merging of two CNs~\cite{Evermann2000CNC}. The reduced CNC alignment scheme proposed here uses a reference one-best sequence rather than a CN as the pivot, in order to tag CN arcs against a reference sequence. A soft loss of aligning reference word $\omega_\tau$ with the $t^{\text{th}}$ CN bin is used
\begin{equation}
    \ell_{t}(\omega_{\tau}) = 1 - P_{t}(\omega_{\tau})
    \label{eqn:loss}
\end{equation}
where $P_t(\omega)$ is a word posterior probability distribution associated with the CN bin at time $t$. The optimal alignment is then found by minimising the above loss.

The extension of the Levenshtein algorithm to lattices, though possible, is computationally expensive ~\cite{Xu2011MinimumBR}. Therefore approximate schemes are normally used~\cite{Povey2002MinimumPE}. Common to those schemes is the use of information about the overlap of lattice arcs and time-aligned reference words to compute the loss
\begin{equation}
    o_{t,\tau} = \max\bigg\{0,\frac{|\min\{e_{\tau}^{*},e_{t}\}| - |\max\{s_{\tau}^{*},s_{t}\}|}{|\max\{e_{\tau}^{*},e_{t}\}|-|\min\{s_{\tau}^{*},s_{t}\}|}\bigg\}
    \label{eqn:overlap}
\end{equation}
where $\{s_t, e_t\}$ and $\{s^{*}_{\tau}, e^{*}_{\tau}\}$ are start and end times of lattice arcs and time-aligned words respectively. In order to yield ``hard'' $0$ or $1$ loss a threshold can be set either on the loss or the amount of overlap.

\presec\presec\presec
\section{Experiments}
\postsec
\label{sec:exp}
Evaluation was conducted on IARPA Babel Georgian full language pack (FLP). The FLP contains approximately 40 hours of conversational telephone speech (CTS) for training and 10 hours for development. The lexicon was obtained using the automatic approach described in \cite{Gales2015aS}. The automatic speech recognition (ASR) system combines 4 diverse acoustic models in a single recognition run \cite{Wang2015aS}. The diversity is obtained through the use of different model types, a tandem and a hybrid, and features, multi-lingual bottlenecks extracted by IBM and RWTH Aachen from 28 languages. The language model is a simple $n$-gram estimated on acoustic transcripts and web data. As a part of a larger consortium, this ASR system took part in the IARPA OpenKWS 2016 competition \cite{Ragni2017aS}. The development data was used to assess the accuracy of confidence estimation approaches. The data was split with a ratio of $8:1:1$ into training, validation and test sets. The ASR system was used to produce lattices. Confusion networks were obtained from lattices using consensus decoding~\cite{Mangu2000FindingCI}. The word error rates of the 1-best sequences are 39.9\% for lattices and 38.5\% for confusion networks. 

The input features for the standard bi-directional recurrent neural network (BiRNN) and CN-based (BiCNRNN) are decision tree mapped posterior, duration and a 50-dimensional fastText word embedding~\cite{Bojanowski2016EnrichingWV} estimated from web data. The lattice-based BiRNN (BiLatRNN) makes additional use of acoustic and language model scores. All forms of BiRNNs contain one $[\overrightarrow{128},\overleftarrow{128}]$ dimensional bi-directional LSTM layer and one $128$ dimensional feed-forward hidden layer. The implementation uses PyTorch library and is available online\footnote{\url{https://github.com/qiujiali/lattice_rnn}}. For efficient training, model parameters are updated using Hogwild! stochastic gradient descent~\cite{Recht2011HogwildAL}, which allows asynchronous update on multiple CPU cores in parallel. 

\tab{exp_1_best} shows the NCE and AUC performance of confidence estimation schemes on 1-best hypotheses extracted from CNs. As expected, ``raw'' posterior probabilities yield poor NCE results although AUC performance is high. The decision tree, as expected, improves NCE and does not affect AUC due to the monotonicity of the mapping. The BiRNN yields gains over the simple decision tree, which is consistent with the previous work in the area~\cite{DelAgua2018SpeakerAdaptedCM, Ragni2018ConfidenceEA}. 
\begin{table}[htbp]
    \centering
    \begin{tabular}{l|cc}
        \toprule
        Estimator & NCE & AUC\\
        \midrule
        1-best CN posteriors       & -0.1978 & 0.9081\\
        \;\;+decision tree   &  0.2755 & 0.9081\\
        \;\;\;\;+BiRNN &  \bf{0.2947} & \bf{0.9197}\\
        \bottomrule
    \end{tabular}
    \caption{Confidence estimation performance on 1-best CN arcs}
    \vspace{-2em}
    \label{tab:exp_1_best}
\end{table}

The next experiment examines the extension of BiRNNs to confusion networks. The BiCNRNN uses a similar model topology, merges incoming arcs using the attention mechanism described in \sect{recursive} and uses the Levenshtein algorithm with loss given by \eqn{loss} to obtain reference confidence values. The model parameters are estimated by minimising average binary cross-entropy loss on all CN arcs. The performance is evaluated over all CN arcs. When transitioning from 1-best arcs to all CN arcs the AUC performance is expected to drop due to an increase in the Bayes risk. \tab{exp_cn} shows that BiCNRNN yields gains similar to BiRNN in \tab{exp_1_best}. 
\begin{table}[htbp]
    \centering
    \begin{tabular}{l|cc}
        \toprule
        Estimator & NCE & AUC\\
        \midrule
        all CN posteriors       & 0.3105 & 0.8243\\
        \;\;+decision tree   & 0.4659 & 0.8243\\
        \;\;\;\;+BiCNRNN & \bf{0.4970} & \bf{0.8365}\\
        \bottomrule
    \end{tabular}
    \caption{Confidence estimation performance on all CN arcs}
    \label{tab:exp_cn}
    \vspace{-1em}
\end{table}

As mentioned in \sect{recursive} there are alternatives to attention for merging incoming arcs. \tab{exp_merge} shows that mean and normalised posterior weights may provide a competitive alternative.\footnote{With lattices, the attention mechanism outperforms other arc merging methods more significantly, which is reported in \tab{exp_recursive}.} 
\begin{table}[htbp]
    \centering
    \begin{tabular}{c|cc}
        \toprule
        Merge & NCE & AUC\\
        \midrule
        $\max$       & 0.4933 & 0.8350\\
        mean         & 0.4966 & 0.8364\\
        normalised posterior & 0.4969 & 0.8363\\
        attention & \bf{0.4970} & \bf{0.8365}\\
        \bottomrule
    \end{tabular}
    \caption{Comparison of BiCNRNN arc merging mechanisms}
    \label{tab:exp_merge}
    \vspace{-1em}
\end{table}

Extending BiRNNs to lattices requires making a choice of a loss function and a method of setting reference values to lattice arcs. A simple global threshold on the amount of overlap between reference time-aligned words and lattice arcs is adopted to tag arcs. This scheme yields a false negative rate of 2.2\% and false positive rate of 0.9\% on 1-best CN arcs and 1.4\% and 0.7\% on 1-best lattice arcs. \tab{exp_loss} shows the impact of using approximate loss in training the BiCNRNN. The results suggest that the mismatch between training and testing criteria, \ie approximate in training and Levenshtein in testing, could play a significant role on BiLatRNN performance. Using this approximate scheme, a BiLatRNN was trained on lattices. 
\begin{table}[htbp]
    \centering
    \begin{tabular}{c|cc}
        \toprule
        Method & NCE & AUC\\
        \midrule
        Levenshtein & \bf{0.4970} & \bf{0.8365} \\
        approximate & 0.4873 & 0.8321 \\
        \bottomrule
    \end{tabular}
    \caption{Comparison of BiCNRNN arc tagging schemes}
    \label{tab:exp_loss}
    \vspace{-0.5em}
\end{table}

\tab{exp_recursive} compares BiLatRNN performance to ``raw'' posteriors and decision trees. As expected, lower AUC performances are observed due to higher Bayes risk in lattices compared to CNs. The ``raw'' posteriors offer poor confidence estimates as can be seen from the large negative NCE and low AUC. The decision tree yields significant gains in NCE and no change in AUC performance. Note that the AUC for a random classifier on this data is 0.2466. The BiLatRNN yields very large gains in both NCE and AUC performance. 
\begin{table}[htbp]
    \centering
    \begin{tabular}{l|cc}
        \toprule
        Estimator & NCE & AUC\\
        \midrule
        all lattice arc posteriors       & -5.0386 & 0.2251\\
        \;\;+decision tree   & -0.0889 & 0.2251\\
        \;\;\;\;+BiLatRNN (post) & 0.3880 & 0.7507 \\
        \;\;\;\;+BiLatRNN (attn) & \bf{0.3921} & \bf{0.7537}\\
        \bottomrule
    \end{tabular}
    \caption{Confidence estimation performance on all lattice arcs}
    \label{tab:exp_recursive}
    \vspace{-2em}
\end{table}

\begin{figure}[ht]
    \centering
    \begin{subfigure}[t]{0.465\linewidth}
        \centering
        \includegraphics[width=1.0\linewidth]{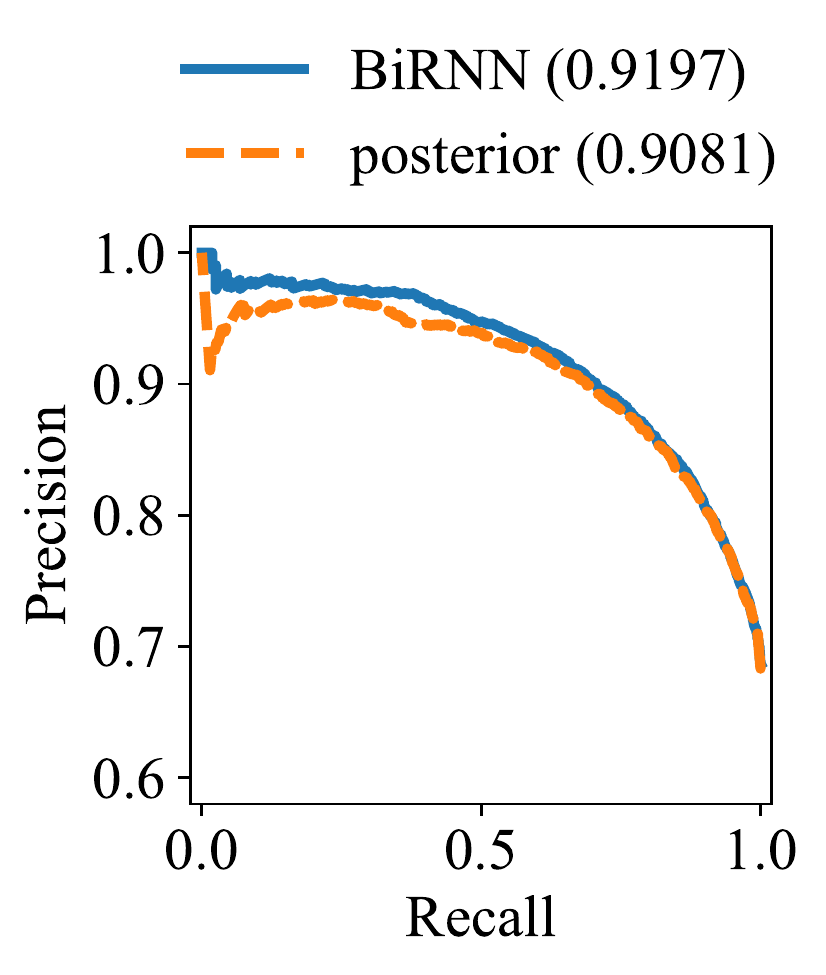}
        \caption{1-best CN arcs}
        \label{fig:pr_onebest}
    \end{subfigure}
    ~ 
    \begin{subfigure}[t]{0.48\linewidth}
        \centering
        \includegraphics[width=1.0\linewidth]{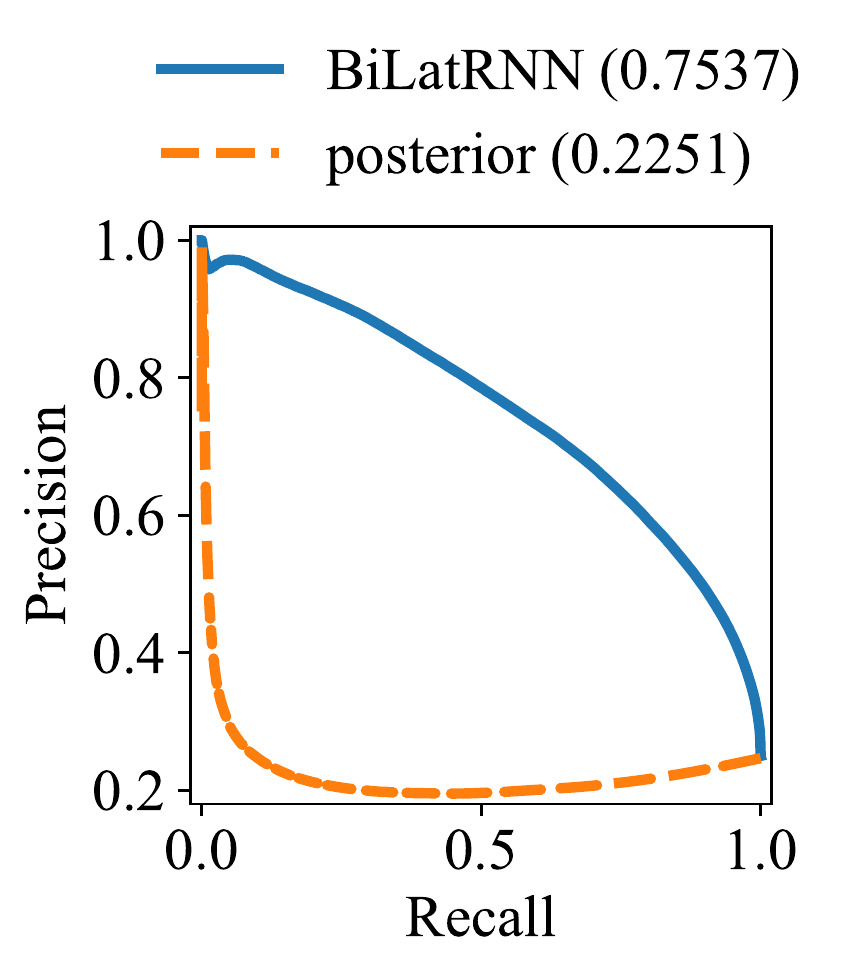}
        \caption{all lattice arcs}
        \label{fig:pr_lattice}
    \end{subfigure}
    \caption{Precision-recall curves for \tab{exp_1_best} and \tab{exp_recursive}}
    \label{fig:pr}
    \vspace{-2em}
\end{figure}

As mentioned in \sect{intro}, applications such as language learning and information retrieval rely on confidence scores to give high-precision feedback~\cite{Knill2018aS} or high-recall retrieval~\cite{Gales2017aL, Ragni2018AutomaticSR}. Therefore, \fig{pr} shows precision-recall curves for BiRNN in \tab{exp_1_best} and BiLatRNN in \tab{exp_recursive}. \fig{pr_onebest} shows that the BiRNN yields largest gain in the region of high precision and low recall which is useful for feedback-like applications. Whereas the BiLatRNN in \fig{pr_lattice} can be seen to significantly improve precision in the high recall region, which is useful for some retrieval tasks. 

\presec
\section{Conclusions}
\postsec
\label{sec:conclusion}
Confidence scores play an important role in many applications of spoken language technology. The standard form of confidence scores are decision tree mapped word posterior probabilities. A number of approaches have been proposed to improve confidence estimation, such as bi-directional recurrent neural networks (BiRNN). BiRNNs, however, can predict confidences of sequences only, which limits their more general application to 1-best hypotheses. This paper extends BiRNNs to confusion network (CN) and lattice structures. In particular, it proposes to use an attention mechanism to combine variable number of incoming arcs, shows how recursions are linked to the standard forward-backward algorithm and describes how to tag CN and lattice arcs with reference confidence values. Experiments were performed on a challenging limited resource IARPA Babel Georgian pack and shows that the extended forms of BiRNNs yield significant gains in confidence estimation accuracy over all arcs in CNs and lattices. Many related applications like information retrieval, speaker adaptation, keyword spotting and semi-supervised training will benefit from the improved confidence measure.

\bibliographystyle{IEEEbib}
\bibliography{main}

\end{document}